\documentclass[aps,nofootinbib,twocolumn,superscriptaddress]{revtex4-1} 
\usepackage{graphicx}

\usepackage[english]{babel}
\usepackage{amssymb,amsfonts,psfrag,amsmath,bm,bbm,indentfirst,subfigure,bbold}
\usepackage{color}
\usepackage{ulem}
\usepackage{natbib}

\RequirePackage[
   hyperindex,colorlinks,bookmarksnumbered,
   plainpages=true,pdfstartview=FitH]{hyperref}
\hypersetup{linkcolor=blue,urlcolor=blue,citecolor=blue}
\usepackage{hyperref}
\usepackage{breakurl}

\definecolor{darkgreen}{rgb}{0,0.5,0}
\definecolor{purple}{rgb}{0.35,0,0.35}
\definecolor{orange}{rgb}{1,0.5,0}
\definecolor{darkred}{rgb}{.7,0,0}
\definecolor{darkblue}{rgb}{0.1,0.1,.6}
\definecolor{grey}{rgb}{.6,.6,.6}
\definecolor{dimgreen}{rgb}{0.2,0.6,0.1}
\definecolor{RoyalBlue}{cmyk}{0.94,0.539,0,0}
\definecolor{DGLorange}{cmyk}{.22,1,1,.2}

\renewcommand{\emph}[1]{\textit{#1}}
\begin{document}
\allowdisplaybreaks

\title{Functional Renormalization Group treatment of the 0.7-analog in quantum point contacts}

\author{Lukas Weidinger}
\author{Christian Schmauder}
\author{Dennis H. Schimmel}
\author{Jan von Delft}

\affiliation{Arnold Sommerfeld Center for Theoretical Physics and Center for
NanoScience,
Ludwig-Maximilians-Universit\"at
M\"unchen, Theresienstrasse 37, D-80333 M\"unchen, Germany}

\date{\today}

\begin{abstract}
We use a recently developed fRG method (extendend Coupled-Ladder Approximation) to study the 0.7-analog in quantum point contacts, arising at the crossing of the 1st and 2nd band at sufficiently high magnetic fields.
We reproduce the main features of the experimentally observed magnetic field dependence of the conductance at the 0.7-analog, using a QPC model with two bands and short-range interactions. In particular, we reproduce the fact that this dependence is qualitatively different, depending on whether the analog is approached from higher or lower magnetic fields. We show that this effect can be explained qualitatively within a simple Hartree picture for the influence of the lowest electrons.
\end{abstract}

\maketitle

\section{Introduction}

In quasi-one-dimensional structures, such as quantum wires or quantum point contacts (QPCs), an in-plane magnetic field induces a Zeeman splitting of different spin subbands. When this splitting equals the one-dimensional level spacing introduced by the lateral confinement of the structure, one finds at the crossings features similar to the 0.7-anomaly, as observed at zero magnetic field. Therefore, these features are called 0.7-analogs \cite{Graham2003}.
The apparent similarities have intertwined the explanation attempts of 0.7-anomaly and 0.7-analogs, prominently featuring spontaneous spin-polarization \cite{Berggren2005}, and quasi-localized states \cite{Meir2008}. 

\begin{figure}
   \includegraphics[scale=0.7]{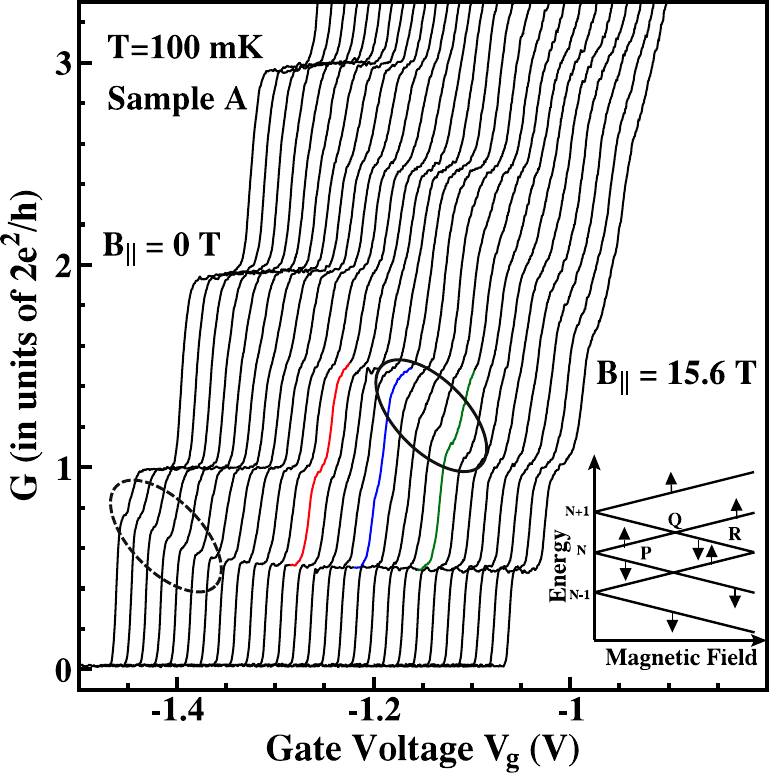} 
   \caption{\small  Fig.~$1$ of \cite{Graham2003} with some additional annotations. As guide for the eye, we colored three curves: The analog of the zero-field conductance step (blue) as well as two curves at magnetic fields $\Delta B=\pm 1.8T$ above (green) and below (red) of the analog of the zero-field conductance step. The 0.7-anomaly is indicated by the dashed ellipse, that of its analog by the solid one. Clearly this 0.7-like behavior is only present if the analog is approached from above. }
\label{graham_modified}
\end{figure}

However, despite observed similarities, there are also features specific to the 0.7-analog that have no counterpart for the 0.7-anomaly. A striking example is the asymmetry in the magnetic field dependence of the conductance, depending on whether the analog is approached from higher or lower fields, see Fig.~\ref{graham_modified}, which is a annotated version of Fig.~$1$ in \cite{Graham2003}. While the 0.7-analog resembles the 0.7-anomaly at higher magnetic fields (green curve), the conductance curves at lower fields (red curve) are much more symmetric and show no sign of a 0.7-shoulder. We used a solid ellipse to mark the higher-field part of the analog, which resembles the 0.7-anomaly (dashed ellipse). 

Some years ago, an interpretation of the 0.7-anomaly was introduced in \cite{Bauer2013} that traces its origins back to the structure of the non-interacting van Hove ridge in the local density of states. This interpretation has been supported by direct conductance calculations of the QPC via the functional renormalization group (fRG).
Following this approach, we use here a recently developed extended Coupled-Ladder Approximation (eCLA) fRG scheme \cite{Weidinger2017} to study the features of the 0.7-analog at the crossing of the $1{\uparrow}$ and $2{\downarrow}$ spin subbands of a QPC, working out the similarities and differences between 0.7-analog and 0.7-anomaly.

We argue that the 0.7-analog physics can be explained in a similar manner as the 0.7-anomaly, evoking a smeared van Hove singularity in the local density of states. However, the effects of the electrons in the lowest spin subband are of critical importance. We demonstrate that these electrons cause the above-mentioned asymmetry in the magnetic field dependence of the conductance and study its dependence on the ratio of intra- to interband interaction strength.

\section{Theoretical model and method}

\subsection{Model}

Since our goal is a qualitative understanding of the 0.7-analog physics, we use here the simplest model that should be able to give us the relevant features. We model the lowest two bands of the QPC via one-dimensional spinful tight-binding chains with an intra- and interband short-ranged interaction. The external magnetic field is modelled by a Zeeman term, splitting the energies of spin up and spin down electrons.
We point out that in experiments one observes additionally to the Zeeman effect also a diamagnetic shift with increasing magnetic field. This shift is understood analytically \cite{Stern1968}, and is expected not to be relevant for the qualitative physics of interest here \cite{Graham2003}. Therefore, we will omit this effect in the present qualitative study, and concentrate on the physics caused by the interactions.    
Our Hamiltonian will thus be of the form:
\begin{align}
\label{Hamiltonian}
H &= - \tau \sum_{i,s,\sigma} \Big[ c^\dagger_{is \sigma} c^{\phantom{\dagger}}_{i+1s \sigma} + h.c. \Big]
   + \sum_{i,s, \sigma} V_{is \sigma} n_{is \sigma} \nonumber \\
  &+ \sum_{i,s} U^{\text{intra}}_{is} n_{i s \uparrow} n_{i s \downarrow}
   + \sum_{i, \sigma_1, \sigma_2} U^{\text{inter}}_i n_{i 1 \sigma_1} n_{i 2 \sigma_2}, 
\end{align}
where $c_{is \sigma}$ annihilates an electron at site $i$ in band $s$ with spin $\sigma$, and $n_{is \sigma} = c^\dagger_{is \sigma} c^{\phantom{\dagger}}_{is \sigma}$ is the corresponding number operator. In our calculations we will use the hopping amplitude $\tau$ as unit of energy, i.e. we measure the onsite energy, $V_{i s \sigma}$, as well as the intraband interaction, $U^{\text{intra}}_{is}$, and the interband interaction, $U^{\text{inter}}_{i}$, in units of $\tau$. 
Within a central region, $i \in [-N,N]$, we use the following form for the potential term:
\begin{align}
V_{i s \sigma} = V_g \exp \Big[ \frac{(i/N)^2}  {1 - (i/N)^2} \Big] + V^{\text{off}}_s + \sigma \frac{B}{2}.   
\end{align}  
Here the first summand leads to a quadratic barrier top in the middle of the QPC with curvature $\Omega_x=2\sqrt{V_g \tau}/N$ and corresponding characteristic length $l_x=a \sqrt{\tau/\Omega_x}$. The second term constitutes the band offset (we choose $V^{\text{off}}_1 = 0$, and therefore use the abbreviation $V^\text{off}:= V^\text{off}_2$) and the third term is the Zeeman splitting.  
Analogous to \cite{Bauer2013}, we take both $V_g$ as well as $U^{\text{intra}}_{is}$, and $U^{\text{inter}}_{i}$ to be zero outside of the central region, where we thus have two non-interacting tight binding leads with the site independent energy offset
\begin{align}
V_{i s \sigma} = V^{\text{off}}_s + \sigma \frac{B}{2}.   
\end{align}
Those can be integrated analytically and their contribution absorbed in the self-energy $\Sigma$ of the central region. Note that this contribution will, however, depend on $V^{\text{off}}_s$, as well as $B$. 
The short-ranged interactions $U^{\text{intra}}_{is}$ and $U^{\text{inter}}_{i}$ are treated as free parameters, chosen as site independent within the middle of the central region, and reduced smoothly to zero at its edges.
All our calculations will be carried out in thermal equilibrium at zero temperature, implying that all states below the chemical potential $\mu$ are filled, all states above are empty.
Our typical observable will be the linear response conductance through the system, and its dependence on the chemical potential $\mu$, as well as on the magnetic field $B$.    

Note that to keep things simple and clear, we have made here several simplifying assumptions. We omit any hopping terms between the two bands, keep the offset between the bands a site independent constant throughout the whole system (in particular the barrier curvature for both bands is the same) and omit any longer-ranged interactions. Furthermore, in all our calculations we will keep $V_g$ constant and vary $\mu$ instead. In terms of the Fermi energy on the central site, $\epsilon_F = \mu - V_g$, this is the same as varying $V_g$ with constant $\mu$, but has the advantage that the bare curvature $\Omega_x (V_g)$ of the barrier does not change.

\subsection{Method}
\begin{figure}
   \includegraphics[scale=0.25]{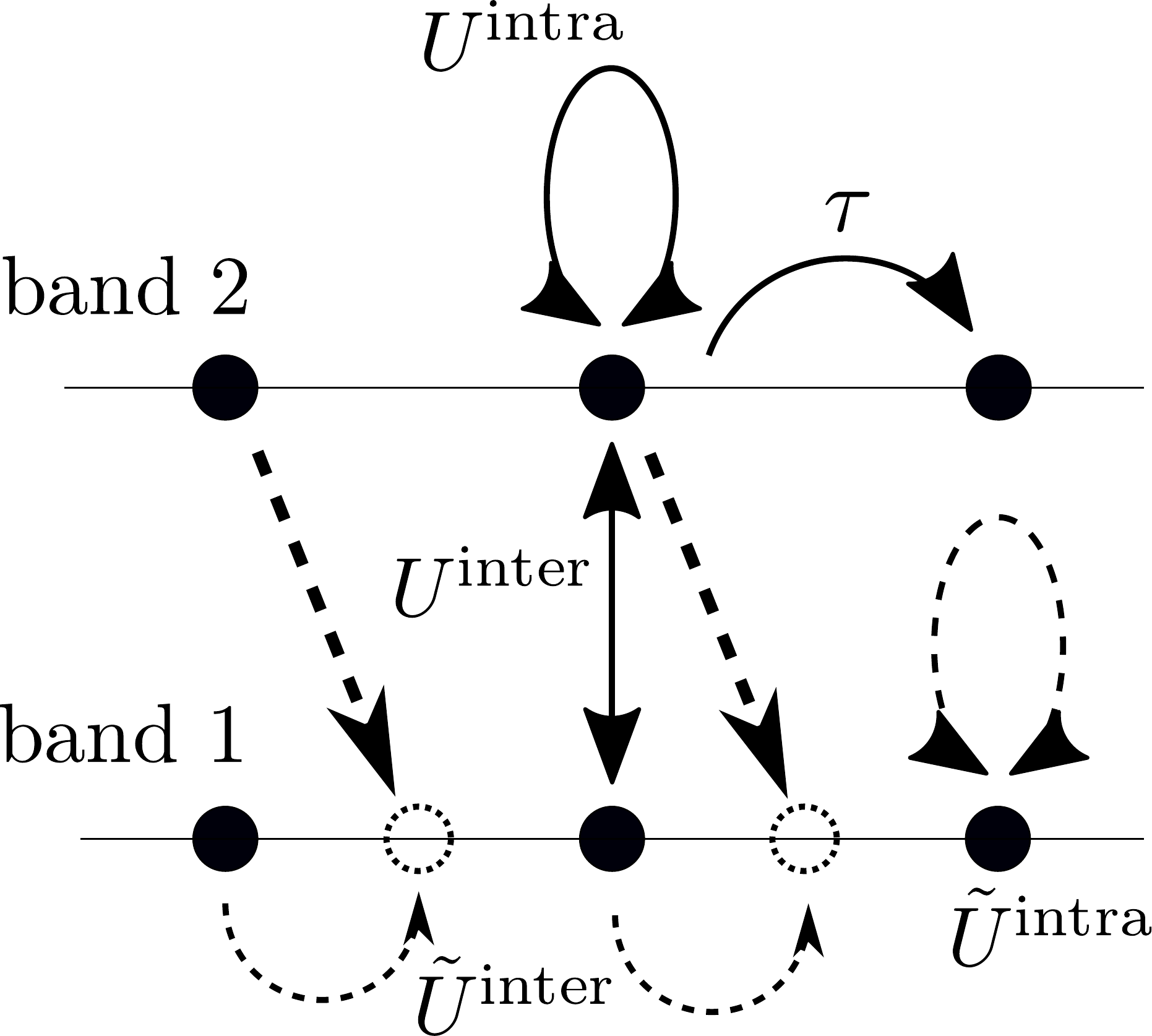} 
   \caption{\small Schematic procedure of interleaving the two bands. Note that in the effective chain, we get again onsite, as well as anisotropic nearest neighbor interactions.}
\label{Chains_interleaved}
\end{figure}

To determine the interaction-induced self-energy, $\Sigma$, and two-particle vertex, $\gamma$, we use the recently introduced eCLA fRG scheme \cite{Weidinger2017} within a static implementation. This scheme was originally designed to treat longer-ranged interactions. It enables the treatment of our two-band model, since it is possible to map the Hamiltonian \eqref{Hamiltonian} onto a one-dimensional chain model with longer-ranged interactions.
For this we simply interleave the different bands, as sketched in Fig.~\ref{Chains_interleaved}, leading to a new effective one-dimensional Hamiltonian, containing interactions between neighboring sites: 
\begin{align}
\label{Effective_hamiltonian}
H_{\text{eff}} &= - \tau \sum_{j, \sigma} \Big[ c^\dagger_{j \sigma} c^{\phantom{\dagger}}_{j+2 \sigma} + h.c. \Big]
                + \sum_{j, \sigma} \tilde{V}_{j \sigma} n_{j \sigma} \nonumber \\
               &+ \sum_{j} \tilde{U}^{\text{intra}}_j n_{j \uparrow} n_{j \downarrow} 
                + \sum_{j, \sigma_1, \sigma_2} \tilde{U}^{\text{inter}}_{2j} n_{2j \sigma_1} n_{2j+1 \sigma_2}.
\end{align}        
Here the new index is given by $j = 2i + s-1$ ($s=1$ is band 1, $s=2$ is band 2), and the coefficients are $\tilde{V}_{j \sigma} = V_{i s \sigma}$, $\tilde{U}^{\text{intra}}_j = U^{\text{intra}}_{is}$, and $\tilde{U}^{\text{inter}}_{j} = U^{\text{inter}}_{i}$.

This Hamiltonian is now in a form suitable for the eCLA approach. Without going into details, we just point out that this method depends crucially on a dimensionless parameter, $L$, called the feedback length in \cite{Weidinger2017}, which determines the spatial extent of the renormalized vertex, $\gamma$. This $L$ has to be chosen large enough to reach convergence, and we will comment on the convergence properties in the beginning of the next chapter.

Finally, the calculation of the linear response conductance, $g=\frac{h}{2e^2} \frac{\partial I}{\partial V}$, from the self-energy and vertices obtained with our fRG method, is carried out via the $T=0$ formula \cite{Datta1997}:
\begin{align}
 g=\frac{1}{2} \sum_{\sigma, s} \left| 2\pi \rho^{\sigma s}(\mu + i0^+) \mathcal{G}_{-NN}^{\sigma s} (\mu + i0^+) \right|^2 ,
\end{align}
where $\rho^{\sigma s}$ is the density of states on the first lead site for spin $\sigma$ and band $s$, and $\mathcal{G}_{-NN}^{\sigma s}$ is the propagator for a electron in band $s$ with spin $\sigma$ from the leftmost to the rightmost site of the central region.

\section{Results}
We use the following general settings in this section: The band offset is chosen as $V^{\text{off}} = 0.1\tau$ and $N=30$, therefore the total number of spatial sites in the central region is $N_\text{tot}= 61$ and correspondingly the total number of effective sites in \eqref{Effective_hamiltonian} is $N_\text{eff-tot}=122$. Furthermore, except for Fig.~\ref{Magnetic_field_convergence_in_L}, we set $V_g=0.5\tau$, implying a curvature $\Omega_x \approx 0.05\tau$. 

In Fig.~\ref{Convergence_in_L}(a) we show the non-interacting, as well as the fully $L$-converged conductance for our two-band model, with the simplest nontrivial interaction configuration, $U^{\text{intra}}_{is} = U^{\text{inter}}_{i} = 0.7\tau$. These values correspond to a typical value for the onsite interactions in a one-band QPC used in \cite{Bauer2013}. The main changes caused by the interaction are the slightly more asymmetric shape of the conductance steps, and the shift to larger chemical potentials observed for the second step. Qualitatively, this shift is caused by the additional interaction energy between the electrons of the two bands (Hartree shift).   

\subsection{Convergence in $L$}

\begin{figure}
   \includegraphics[scale=1.0]{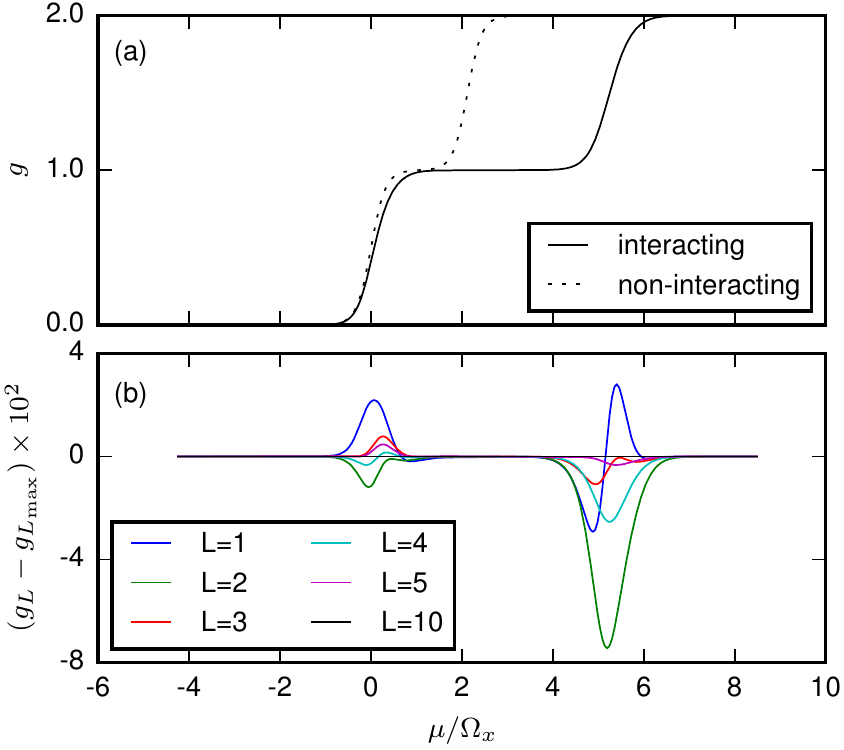} 
   \caption{\small (a) Conductance $g$ for the non- as well as the interacting system for $L=10$ as function of $\mu$. (b) Difference between the conductance for several $L$ values to the converged result (at $L=L_\text{max}=10$). We see that convergence is achieved around $L=5$. }
\label{Convergence_in_L}
\end{figure}

Before we proceed, let us first discuss the convergence of our method with respect to the feedback length $L$. For a one-band QPC with onsite interactions, $L$ has to be of the order of the characteristic length of the harmonic barrier top to achieve convergence: $L \approx l_x/a$, with the lattice spacing $a$. For our interleaved two-band system, we would thus simply expect $L \approx 2 l_x/a$, since the effective distance between two points of the same band is doubled and the effect of the now finite interaction range on the convergence should be negligible, since the introduced nearest neighbor interaction is still much shorter than $l_x$.
In Fig.~\ref{Convergence_in_L}(b) the convergence behavior in $L$ is shown.
We see that the convergence for the two-band model is achieved around $L \approx 5$. Since in our system $l_x \approx 4.6a$, this shows that $L$ can in fact be chosen smaller than the naive guess, $L \approx 2 l_x/a$, indicating stabilizing feedback effects between the two bands.

As a side remark, we point out that the finite extent of the renormalized vertex beyond the lowest value (i.e. $L>1$) is actually important to treat the screening properties between the two bands. This will be seen in the next section when we study the magnetic field dependence of the conductance. 

\subsection{Small magnetic field}

\begin{figure}
   \includegraphics[scale=1.0]{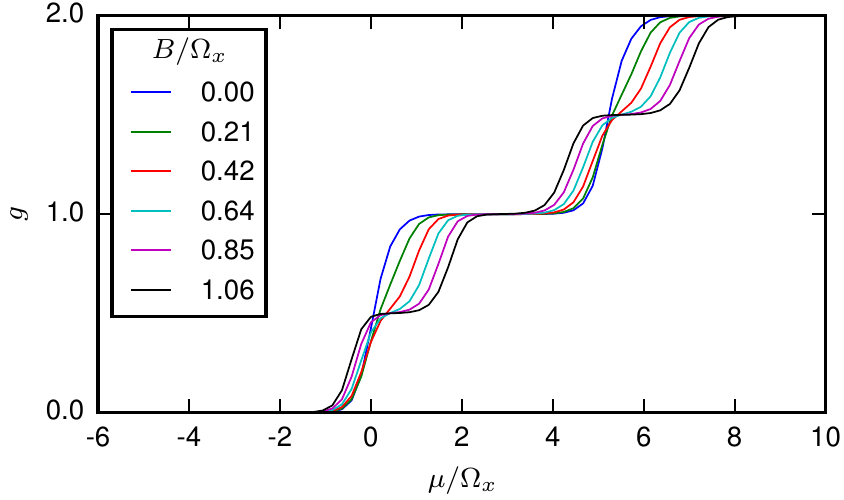} 
   \caption{\small Conductance at low magnetic fields, i.e.\ with $B \ll V^\text{off}=0.1\tau \approx 2.12\,\Omega_x$. The second step is more symmetric and broader than the first step.}
\label{Low_field_conductance}
\end{figure}

Before we look at the 0.7-analog, we want to take a brief look at the properties of the conductance at magnetic fields much smaller than the band spacing, $B \ll V^\text{off}$, see Fig.~\ref{Low_field_conductance}.  

There are two main observations we make here: First, we see that the magnetic field dependence of the second step is more symmetric, indicating that the interaction of the electrons in the second band is screened by electrons in the first band.  
Second, we see that the second conductance step is broader than the first one. This feature can be understood in a simple Hartree picture: While increasing $\mu$ during the second step, electrons are still filling up the lowest band, leading to a increasing Hartree shift for the electrons in the higher band. As a result the second step gets broadened. The effect that the electrons in the first band change the form of the second conductance step is quite generic and will be also encountered in the 0.7-analog case.

\begin{figure}
   \includegraphics[scale=1.0]{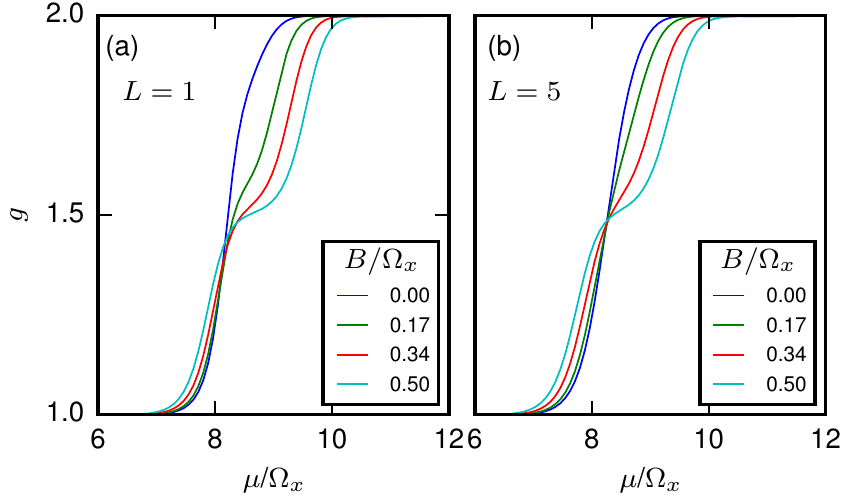} 
   \caption{\small Magnetic field dependence of the second conductance step. (a) $L=1$, (b) $L=5$. We see that the second conductance step is more symmetric in the $L=5$ case, indicating a better screening of interactions in the second band.}
\label{Magnetic_field_convergence_in_L}
\end{figure}

Here it is also interesting to look at the $L$-dependence of the conductance with various magnetic fields. Particularly for longer QPCs, where $l_x \gtrsim 5a$, the increase in $L$ has a visible impact, see Fig.~\ref{Magnetic_field_convergence_in_L}.   
For $L=1$ the second conductance step is very asymmetric, but becomes more symmetric with increasing $L$, due to the screening of the interaction in the second band instigated by electrons in the first band. 
The curvature, $\Omega_x \approx 0.03\tau$ ($V_g=0.2\tau$), which we used here is comparable to the one in a previous fRG study \cite{Eissing2013} of the two-band model. However, in that work the results were not converged in $L$, therefore underestimating screening effects.

\subsection{0.7-analog at large magnetic field}
\label{07analog_at_large_field}

Having studied the properties of the two-band model at low magnetic fields, we are now prepared to tackle the 0.7-analog. This analog appears at the crossing of the $1{\uparrow}$ and the $2{\downarrow}$ spin subbands at a magnetic field, $B = B_c$, which is of the order of the energy separation of the two bands $V^\text{off}$ (determined by the confinement in the lateral direction). This situation resembles the situation given in the 0.7-anomaly, in the sense that two particle species are competing while trying to get through the QPC. Therefore, one might naively expect that the 0.7-analog shows the same features as the 0.7-anomaly. However, this is only partially true.  
While for $\Delta B = B-B_c > 0$ the experimentally measured conductance shows the typical feature of 0.7-physics, namely the development of a shoulder with increasing magnetic field, this feature is missing for $\Delta B<0$.    

\begin{figure}
   \includegraphics[scale=1.0]{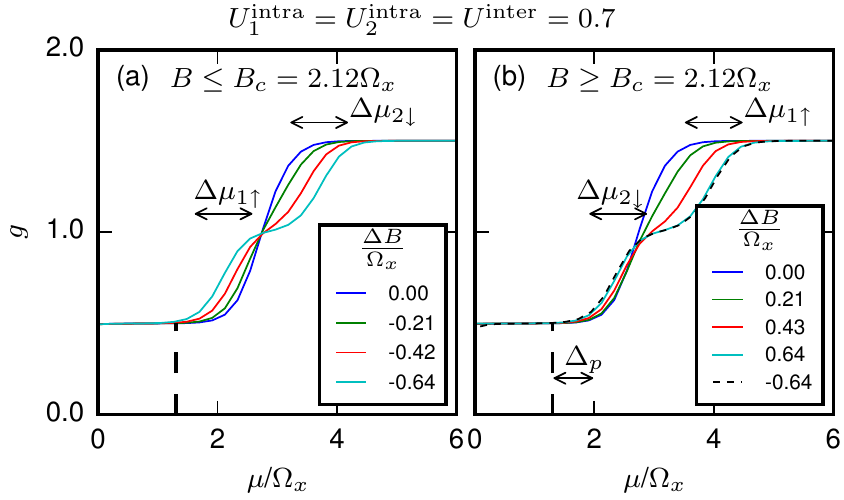} 
   \caption{\small (a) Conductance curves for $B \leq B_c=0.1\tau \approx 2.12\,\Omega_x$ and (b) for $B \geq B_c$ at equal intra- and interband interaction strengths. In (b) the dotted curve is the manually shifted curve, $\Delta B/\Omega_x=-0.64$, from (a). We see that it has exactly the same form as the corresponding curve for $\Delta B/\Omega_x=0.64$. The quantities $\Delta \mu_{1\uparrow}$, $\Delta \mu_{2\downarrow}$ measure the width of the corresponding half-steps and $\Delta_p$ indicates the pinch-off shift between $\Delta B<0$ and $\Delta B>0$, see also Fig.~\ref{Combined_explanation}. }
\label{Analog_equal_interactions}
\vspace{-0.2cm}
\end{figure}

In trying to understand the underlying physics, we first start with the simplest interaction model, $U^{\text{intra}}_{is} = U^{\text{inter}}_{i} \equiv U = 0.7\tau$, which we already used in the last sections. Fig.~\ref{Analog_equal_interactions} shows the resulting conductance.

We make two main observations:  First, the curves for $\Delta B<0$ lie approximately symmetrically around the $B_c$ curve, while the $\Delta B > 0$ curves do not. However, second, the actual \textit{shapes} of corresponding curves, i.e.\ for $B = B_c - \Delta B$ and $B = B_c + \Delta B$, are very similar, they are just offset by different amounts.    

This behavior can be understood by a similar argument as used for the broadening of the conductance step in the low magnetic field case above. As already mentioned, in a case with only $1{\uparrow}$ and $2{\downarrow}$ particles, the situation would be completely symmetric. Therefore the different behavior
must stem from the other particles in the system. 
Since in the analog case the $2{\uparrow}$ spin subband lies much higher than the chemical potential and is therefore empty, the $1{\downarrow}$ particles must be responsible for the change of situation. 

\begin{figure*}
  \includegraphics[width=183mm]{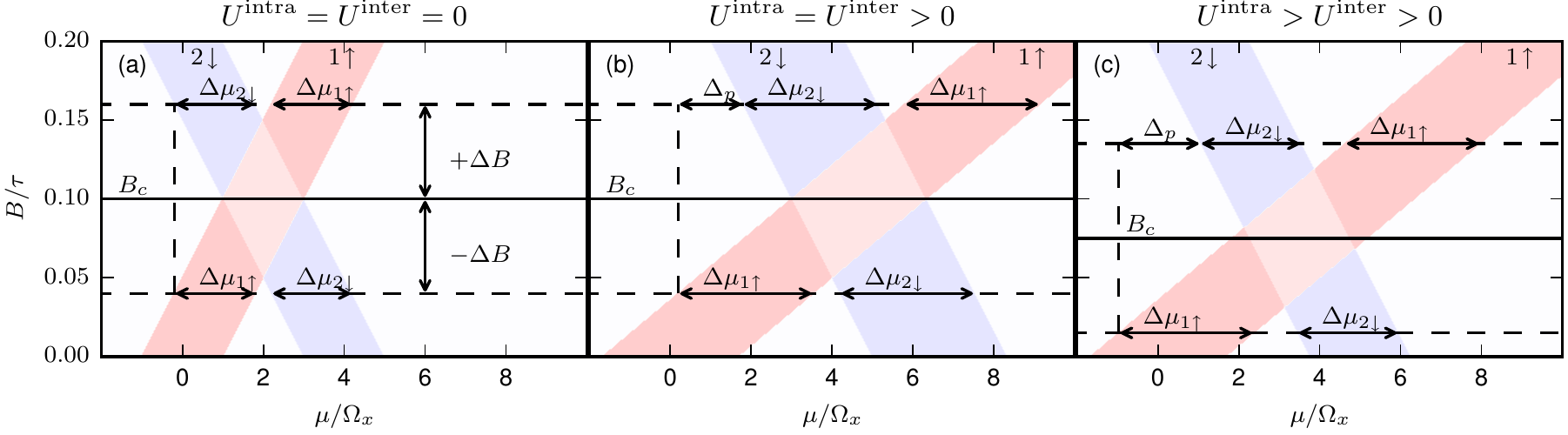} 
  \caption{\small Schematic behavior of the Hartree renormalized barriers of the $1{\uparrow}$ (red) and $2{\downarrow}$ (blue) particles as function of $\mu$ and $B$. The colored regions indicate where $|b_{a} - \mu| < \Omega_x/2$, i.e. the regions within which the conductance steps occur. (a) Non-interacting case: $B_c = V^\text{off}$, no pinch-off asymmetry, no shape asymmetry. (b) $U^\text{intra} = U^\text{inter}$: $B_c = V^\text{off}$, pinch-off asymmetry ($\Delta_p>0$), no shape asymmetry. (c) $U^\text{intra} > U^\text{inter}$: $B_c < V^\text{off}$, pinch-off asymmetry ($\Delta_p>0$) and shape asymmetry ($\Delta \mu_{1\uparrow} > \Delta \mu_{2\downarrow}$). }
\label{Combined_explanation}
\end{figure*}

Both of our observations can be explained by taking the effect of the $1{\downarrow}$ electrons in a simple Hartree argument into account: 
The Hartree-shift on particle species $a$ induced by the $1{\downarrow}$ particles is given by $ E^a_H = U_a n_{1\downarrow}$, where $n_{1\downarrow}$ is the density of the $1{\downarrow}$ particles and $U_a$ denotes the appropriate interaction ($U^\text{intra}$ for $a=1{\uparrow}$, $U^\text{inter}$ for $a=2{\downarrow}$). Assuming that the chemical potential $\mu$ is already far above the $1{\downarrow}$ van Hove ridge, $\mathcal{A}_{1\downarrow}(\omega) \approx \mathcal{A}_{1\downarrow}$ will be approximately constant, and the Hartree shift will be approximately of the form
$ E^a_H \approx U_a (n^0_{1\downarrow} + (\mu - b_{1\downarrow}) \mathcal{A}_{1\downarrow})$, with a constant $n^0_{1\downarrow}$ and the barrier top of the $1{\downarrow}$ particles given by $b_{1\downarrow} = V_g - \frac{B}{2}$.
Leaving the other interactions aside for a moment, we can readily write down the $\mu$- and $B$-dependence of the renormalized barrier tops of the $1{\uparrow}$ and $2{\downarrow}$ particles:    
\begin{align}
b_{1\uparrow} &=  V_g + \frac{B}{2} + E^{1\uparrow_H}(B,\mu) \\
b_{2\downarrow}   &=  V_g + V^\text{off} - \frac{B}{2} + E^{2\downarrow_H}(B,\mu). 
\label{Barrierpositions}
\end{align}
The qualitative behavior of this equations is shown in Fig.~\ref{Combined_explanation}, and provides a good explanation for the observed phenomena:
In contrast to the non-interacting case (Fig.~\ref{Combined_explanation}(a)), we obtain for $U^\text{intra} = U^\text{inter}$ a pinch-off asymmetry, $\Delta_p$, between the pinch-offs at magnetic fields above and below the analog, see Fig.~\ref{Combined_explanation}(b). Taking into account the interaction between $1{\uparrow}$ and $2{\downarrow}$ (whose main effect is a broadening of the second half-step), this results in the more symmetric arrangement of the two half-steps around the crossing curve for $\Delta B <0$, and to a more asymmetric situation in the $\Delta B > 0$ case. However, we see that the shape of corresponding curves is the same since the $\mu$-width of the half-steps, $\Delta \mu_{1\uparrow}$ and $\Delta \mu_{2\downarrow}$, is equal.

\begin{figure}
   \includegraphics[scale=1.0]{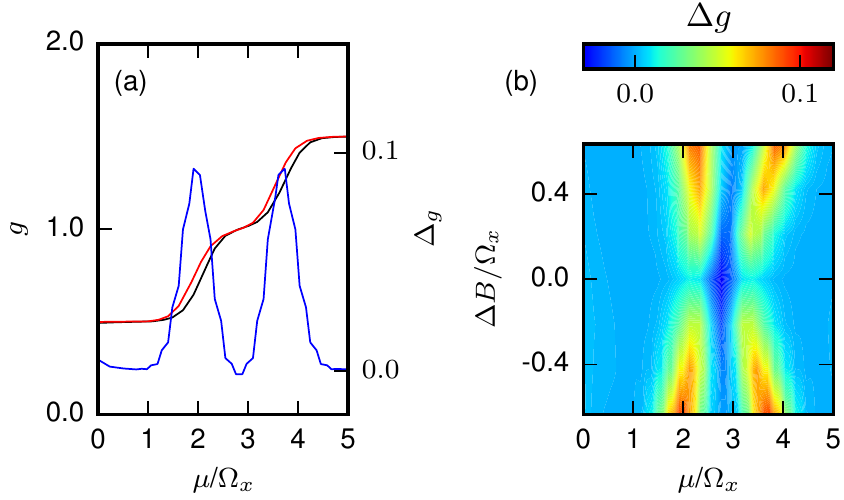} 
   \caption{\small Illustration of the asymmetry in $\mu$. (a) Conductance curve $\Delta B/\Omega_x=-0.64$ from Fig.~\ref{Analog_equal_interactions} (black), together with its mirror image (red) under inversion around the $g=1.0$ point, and the difference $\Delta g$ between the two curves (blue). (b) Colorplot of $\Delta g$ as a function of magnetic field and chemical potential. We see that the strength of the curve asymmetry is symmetric around $B_c$.}
\label{Analog_equal_interactions_show_asymmetry}
\end{figure}

If we compare this to experiment \cite{Graham2003}, we see that this setting reflects only partially the experimental situation: While the half-steps are indeed arranged more symmetrically for $\Delta B <0$ than for the $\Delta B > 0$ case, also the form of the corresponding curves themselves differs substantially in experiment. For $\Delta B>0$, the conductances curves are much more asymmetric in the $\mu$ behavior, developing a 0.7-analog plateau, while for $\Delta B <0$ they are not. To analyze this quantitatively in our calculation, we introduce the `conductance asymmetry` $\Delta g(\mu) = g_m(\mu) - g(\mu)$, where $g_m(\mu)$ is the mirror image of $g(\mu)$ around the point $g(\mu)/g_0 = 1.0$ under reflection in both the horizontal and vertical direction. The more asymmetric the conductance curve is in $\mu$, the larger gets the modulus of $\Delta g$. This is illustrated in Fig.~\ref{Analog_equal_interactions_show_asymmetry}(a). Fig.~\ref{Analog_equal_interactions_show_asymmetry}(b) shows the dependence of this asymmetry $\Delta g$ on the magnetic field. We see that contrary to the experiment the asymmetry is equally strong above and below the crossing value $B_c$.

This indicates that our description up to now lacks an important ingredient. We will argue in the following that this is due to the unphysical choice $U^\text{intra} = U^\text{inter}$. Generically, one would expect $U^\text{inter} < U^\text{intra}_2 < U^\text{intra}_1$. The first statement is due to the smaller overlap of the transversal wave functions between different bands, the second because the transversal wave function in the second band is spread out wider than in the first band. Both effects lead to a weakening of the effective one-dimensional interaction strength. Estimates for the ratios of this different interaction strengths can be obtained in a similar manner as in \cite{Weidinger2017}, see appendix, and yield $U^{\text{intra}}_2/U^{\text{intra}}_1 \approx 0.77$ and $U^{\text{inter}}/U^{\text{intra}}_1 \approx 0.36$. Keeping our previous $U^{\text{intra}}_1$ fixed, this leads approximately to $U^{\text{intra}}_2 =0.5$ and $U^{\text{inter}} =0.3$ .

\begin{figure}
   \includegraphics[scale=1.0]{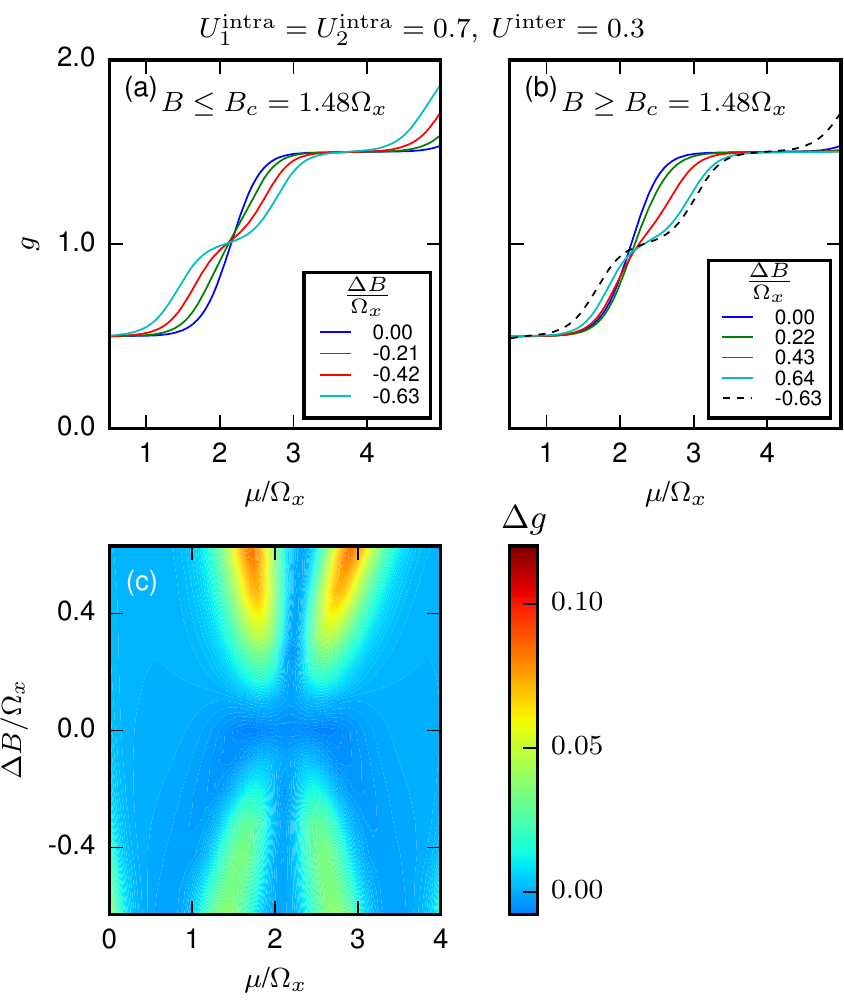} 
   \caption{(a,b) same plot as in Fig.~\ref{Analog_equal_interactions}, but for larger intra- than interband interaction ($U^\text{intra}_{1,2}=0.7\tau$, $U^\text{inter}=0.3\tau$), resulting in $B_c \approx 1.48$. In (b) the dashed curve is again the $\Delta B/\Omega_x=-0.63$ curve from (a), manually shifted such that it intersects the corresponding $\Delta B/\Omega_x=0.63$ curve at the $g/g_0=1$ point. However contrary to Fig.~\ref{Analog_equal_interactions}(b) the shape of the two curves does not coincide. (c) Colorplot of the shape asymmetry. In contrast to Fig.~\ref{Analog_equal_interactions_show_asymmetry} (b), we see that the asymmetry is clearly stronger for $\Delta B>0$ than for $\Delta B<0$. }
\label{Analog_different_interactions}
\end{figure}

To investigate the influence of these differences in interaction strength, we proceed in two steps. 
In the ideal case where the analog region is well separated from the $2{\uparrow}$ conductance step, we expect that the influence of $U^{\text{intra}}_2$ at the analog is not important, since the barrier for the $2{\uparrow}$ electrons is way above the chemical potential. Therefore, we will first keep $U^{\text{intra}}_2$ equal to $U^{\text{intra}}_1=0.7\tau$ and investigate the influence of 
a reduction of  $U^\text{inter}=0.3\tau$ alone.  
In Fig.~\ref{Analog_different_interactions} we show the resulting conductance curves.
Again, we encounter a pinch-off shift of the higher spin subband steps, however due to the different interaction strengths, the crossing point $B_c$ is now shifted, too.  
More importantly, we see that in addition to the pinch-off asymmetry, also the shape of corresponding curves for $\Delta B<0$ and $\Delta B>0$ differ, the curves for $\Delta B<0$ being much more symmetric than the $\Delta B>0$ curves. This is the behavior also observed in experiment and for further reference, we will call it the `shape asymmetry`.

These features can be readily explained with our Hartree picture for the renormalized barrier positions \eqref{Barrierpositions}. Their behavior for $U^\text{intra} > U^\text{inter}$ (i.e. the Hartree shift for the $2{\downarrow}$ subband is smaller than for the $1{\uparrow}$ subband) is shown in Fig.~\ref{Combined_explanation}(c). We see two immediate effects: (i) The $2{\downarrow}$ subband is shifted to lower values of $\mu$ and therefore the value of the magnetic field $B_c$ where the two subbands cross is shifted to lower magnetic fields, as encountered in the Fig.~\ref{Analog_different_interactions}.  
(ii) The width $\Delta \mu_{2\downarrow}$ of the $2{\downarrow}$ half-step is decreased, therefore yielding the shape asymmetry: For $\Delta B<0$ the first half-step ($1{\uparrow}$) is broader than the second half-step ($2{\downarrow}$), thus counteracting the asymmetry introduced by the interband interaction between the competing particles themselves and leading in total to a more symmetric curve. For $\Delta B>0$ the effect is reversed, leading to a more asymmetric curve.

\begin{figure}
   \includegraphics[scale=1.0]{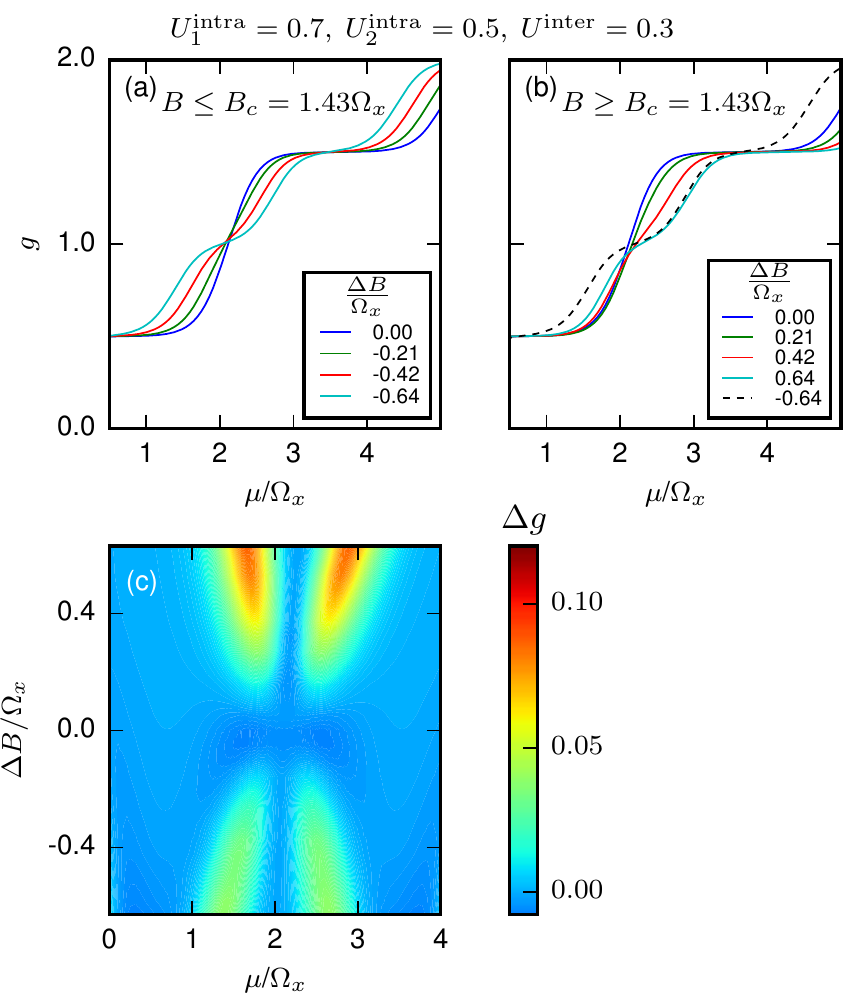} 
   \caption{(a,b) same plot as in Fig.~\ref{Analog_equal_interactions}(a,b) and Fig.~\ref{Analog_different_interactions}(a,b), but for three different interactions: $U^\text{intra}_{1}=0.7\tau$, $U^\text{intra}_{2}=0.5\tau$ and $U^\text{inter}=0.3\tau$. In comparison to Fig.~\ref{Analog_different_interactions}, the crossing point is slightly reduced to $B_c = 1.43 \Omega_x$, however the asymmetry persists. (c) Colorplot of the shape asymmetry, which stays very similar to Fig.~\ref{Analog_different_interactions}(c).}
\label{Analog_different_interactions_different_U_intra}
\end{figure}

Now as a last step, we finally also reduce $U^{\text{intra}}_2 = 0.5\tau < U^{\text{intra}}_1$. The results are shown in Fig.~\ref{Analog_different_interactions_different_U_intra}. We see that the reduction of $U^{\text{intra}}_2$ slightly shifts the crossing point $B_c$ to lower values of the magnetic field, however the shape asymmetry introduced by the lowering of $U^{\text{inter}}$ stays intact. Thus in terms of Fig.~\ref{Combined_explanation}, the net effect of the reduction of $U^{\text{intra}}_2$ is simply a slight shift of the blue $2{\downarrow}$ barrier top position stripe to the left, i.e. to lower values of $\mu$, without changing its slope.     

\subsection{Limitations}
\begin{figure}
   \includegraphics[scale=1.0]{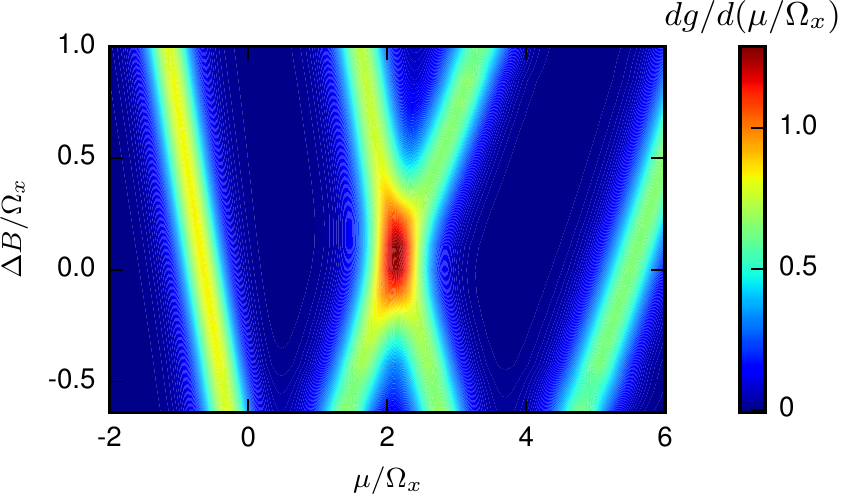} 
   \caption{\small Colorplot of the transconductance as function of $B$ and $\mu$. Note the more pronounced asymmetry at the $\Delta B>0$ than the $\Delta B<0$ part of the crossing region. }
\label{Analog_transconductance}
\end{figure}

A limitation of our static zero temperature calculation is that we have no access to inelastic processes. We suspect that this leads to a main difference between our results and experimental observations, namely that we do not see a pronounced finite temperature plateau in the conductance. This can be clearly seen by comparing the transconductances $dg/d\mu$, see Fig.~\ref{Analog_transconductance}, where we do not observe the `gap` at $\Delta B > 0$ as in the experimental data, cf.\ Fig.~2(a) in \cite{Graham2003}  or  Fig.~1(b) in \cite{Graham2007}. However, we also see in the transconductance, that for $\Delta B>0$ the broadening of the conductance curve in the second half-step is more pronounced than for $\Delta B<0$, where the half-steps are more symmetric in position as well as slope. 

\vspace{0.5cm}

\section{Conclusion}

We have studied the 0.7-analog in QPCs using a two-band model with intra- and interband onsite interactions and found that we could qualitatively reproduce the magnetic field dependence of the conductance around the analog. In particular, we could reproduce the asymmetry in the conductance, depending on whether the analog is approached from higher or lower magnetic fields.

Due to our use of a static fRG scheme, we were not able to investigate finite temperature properties of the analog, which is an interesting direction for further research.  

\begin{acknowledgments}
\vspace{-0.2cm}
We thank Prof.\ Dr.\ D.\ A.\ Ritchie for letting us use Fig.~$1$ from their publication \cite{Graham2003}. Furthermore, we gratefully acknowledge support from the Cluster of Excellence \emph{Nanosystems Initiative Munich}. 
\end{acknowledgments}

\vspace{-0.3cm}
\appendix 
\section*{Appendix: Estimate of the QPC interaction strengths}
Following the approach of \cite{Lunde2009}, we calculated in \cite{Weidinger2017} the intraband interaction for a QPC with a single band that resulted from a screened Coulomb interaction. This was done by taking only the ground state $\phi_1$ of the transversal $y$-direction (in the two-dimensional electron gas plane) into account. Since the confinement in $y$-direction can be approximated by a harmonic potential, $\phi_1$ is simply the ground state of a harmonic oscillator. In a QPC with two bands, we additionally also take the first excited state of the harmonic $y$-confinement into account. The computation of the resulting matrix elements for the interaction between two effective one dimensional states at $x_0$ and $x_1$ can be done analogously to the one-dimensional case and yield in terms of integrals over the relative coordinate $r$ in the transversal direction:
\begin{align}
&U^{\text{intra}}_1(x_0,x_1) = (l_y^2(x_0) + l_y^2(x_1))^{-\tfrac{1}{2}} \int {\rm d}r \, g(r) \\ 
&U^{\text{intra}}_2(x_0,x_1) = (l_y^2(x_0) + l_y^2(x_1))^{-\tfrac{9}{2}} \int {\rm d}r \, g(r) \nonumber \\ 
                            & \times\Big[ 3 l_y^2(x_0) l_y^2(x_1) \left( l_y^2(x_0)+l_y^2(x_1) \right)^2 + \nonumber \\
                            &+ \left(l_y^2(x_0)+l_y^2(x_1)\right) \left( l_y^4(x_0)-4 l_y^2(x_0) l_y^2(x_1) + l_y^4(x_1) \right) r^2 \nonumber \\
                            &+ l_y^2(x_0) l_y^2(x_1) r^4 \Big] \\
&U^{\text{inter}}(x_0,x_1) =  (l_y^2(x_0) + l_y^2(x_1))^{-\tfrac{5}{2}} \int {\rm d}r \, g(r) \nonumber \\
& \left[ l_y^4(x_1)+l_y^2(x_0)(l_y^2(x_1)+r^2)\right],
\end{align} 
where $l_y(x)$ is the ($x$-dependent) characteristic length in $y$-direction, $e$ the electron charge, $\kappa$ the dielectric constant and $g(r)$ (which consists of the screened Coulomb interaction, as well as the lateral confinement) is given by: 
\begin{align}
g(r) &= \frac{e^2}{\kappa} \Big[ \frac{1}{\sqrt{(x_0-x_1)^2+r^2}}-\frac{1}{\sqrt{(x_0-x_1)^2+r^2+l_s^2}}\Big] \nonumber \\
&\times e^{-r^2/(2(l_y^2(x_0)+l_y^2(x_1)))},
\end{align}
where $l_s$ is the screening length. All these contributions are logarithmically divergent for $x_0 \rightarrow x_1$. In this work, we make the simplest approximation and ignore the position dependence of the $U$'s, by setting them to their value in the QPC center. Then we obtain for the ratios of the different effective interaction strengths used in section~\ref{07analog_at_large_field}:   
\begin{align}
\frac{U^{\text{intra}}_2}{U^{\text{intra}}_1} &= \lim_{ x_1 \rightarrow 0} \frac{U^{\text{intra}}_2(0, x_1)}{U^{\text{intra}}_1(0, x_1)} \approx 0.77  \\
\frac{U^{\text{inter}}}{U^{\text{intra}}_1} &= \lim_{ x_1 \rightarrow 0} \frac{U^{\text{inter}}(0, x_1)}{U^{\text{intra}}_1(0, x_1)} \approx 0.36,  
\end{align}
where in the last step we used a ratio $l_s/l_y(0)=3$, which could for example be realized in a QPC with $l_s=50{\rm nm}$ and $l_y=17{\rm nm}$ which corresponds in a GaAs 2DEG to a curvature $\Omega_y = 2{\rm meV}$.  
\bibliography{07_analog_citations}
\end{document}